# Magnetic Josephson Junctions and Superconducting Diodes in Magic Angle Twisted Bilayer Graphene

J. Díez-Mérida[1], A. Díez-Carlón[1], S. Y. Yang[1], Y.-M. Xie[2], X.-J. Gao[2], K. Watanabe[3], T. Taniguchi[3], X. Lu[1], K. T. Law[2] and Dmitri K. Efetov[1]*

1.   ICFO - Institut de Ciencies Fotoniques, The Barcelona Institute of Science and Technology, Castelldefels, Barcelona, 08860, Spain
2.   Department of Physics, Hong Kong University of Science and Technology, Clear Water Bay, Hong Kong, China
3.   National Institute of Material Sciences, 1-1 Namiki, Tsukuba, 305-0044, Japan

*E-mail: dmitri.efetov@icfo.eu

**The simultaneous co-existence and gate-tuneability of the superconducting (SC), magnetic and topological orders in magic angle twisted bilayer graphene (MATBG) open up entirely new possibilities for the creation of complex hybrid Josephson junctions (JJ). Here we report on the creation of gate-defined, magnetic Josephson junctions in MATBG, where the weak link is gate-tuned close to the correlated state at a moiré filling factor of $\nu = -2$. A highly unconventional Fraunhofer pattern emerges, which is phase-shifted and asymmetric with respect to the current and magnetic field directions, and shows a pronounced magnetic hysteresis. Interestingly, our theoretical calculations of the JJ with a valley polarized $\nu = -2$ with orbital magnetization as the weak link explain most of these unconventional features without fine tuning the parameters. While these unconventional Josephson effects persist up to the critical temperature $T_c \sim 3.5K$ of the superconducting state, at temperatures below $T < 800mK$, we observed a pronounced magnetic hysteresis possibly due to further spin-polarization of the $\nu = -2$ state. We demonstrate how the combination of magnetization and its current induced magnetization switching in the MATBG JJ allows us to realize a programmable zero field superconducting diode, which represents a major building block for a new generation of superconducting quantum electronics.**

Electronic coupling between materials with competing ground states can lead to the creation of exotic electronic phases. Of particular interest are hetero-junctions between superconductors, magnets and topological insulators, where especially Josephson junctions (JJ) have attracted formidable attention. Magnetic JJs allow spintronic applications through the creation of spin-filters[1–5], spin-triplet supercurrent[6–8] and $\pi$ junctions (cite)[9–12], whereas topological JJs allow applications in quantum information processing and in lossless electronics, through the creation of $4\pi$ junctions[13–15], superconducting diodes[52,53,54] and Majorana bounds states [16]. One major difficulty in the creation of such JJs lies in the engineering of ultra-clean interfaces between the different material species, which are essential for efficient electronic coupling between the individual phases.

A single, two-dimensional material which would host all of these emergent phases at once, would allow to overcome these issues. It would permit to induce gate-defined, ultra-clean homojunctions between all the different phases, and so open up a new avenue for the creation of a new generation of superconducting electronics. The recently discovered quantum phases in the flat-bands of $\theta_m \sim 1.1°$ magic angle twisted bilayer graphene (MATBG) include correlated insulators (CI)[17–21], superconductors (SC)[19,20,22–24], orbital magnets (OM)[20,25,26] and interaction induced correlated Chern insulators (CCI)[27–31]. Remarkably, due to MATBG's two-dimensionality and ultra-low carrier density, it is possible to use an electrostatic gate to tune between the different phases, and induce reversible transitions between the SC, CCI and OM phases [32]. Although there have already been reports on the creation of MATBG JJs through the use of local gates [33,34], JJs with the supercurrent mediated by the strongly correlated magnetic or topological weak links have so far not been achieved.

Here, we demonstrate the creation of a magnetic JJ in a locally gated MATBG device, when the weak link is set close to half-filling of the hole band. The device consists of a van der Waals (vdW) hetero-structure of graphite/hBN/MATBG/hBN/graphite, as shown in Fig 1a. Metallic graphite layers are capacitively coupled to the MATBG, through an insulating hexagonal boron nitride layers (hBN) of ~ 10 nm thickness. The carrier density in the MATBG sheet $n$ is electrostatically tuned by both top and bottom gates $n = C_{BG}V_{BG} + C_{TG}V_{TG}$, where $(C_{BG}, C_{TG})$ are the respective capacitances and $(V_{BG}, V_{TG})$ gate voltages. In the center of the device, the top graphite layer is separated by a narrow channel of length $d$ = 150 nm, which creates a region in the MATBG that is almost not coupled to the top gate, and whose carrier density is mainly set by the back gate voltage $n_J \sim C_{BG}V_{BG}$. Hence, by applying different values of $V_{BG}$ and $V_{TG}$, it is possible to locally vary the carrier concentration in the channel region, which allows to create gate-defined junctions of length $d_j \sim$ 100 nm in the MATBG, as is confirmed by electrostatic simulations and is highlighted in Fig. 1b.

Fig. 1c shows the 4-terminal longitudinal resistance $R_{xx}$ as a function of $V_{BG}$ ($V_{TG} = 0$ V) at base temperature $T$ = 35 mK, and for different perpendicular magnetic fields $B$. From Hall and quantum magneto-oscillation measurements (see SI) we extract a twist-angle of $\theta$ = 1.11° ± 0.02°. We observe well pronounced CI states, which give rise to peaks of high resistance at integer electron and hole fillings of the moiré unit cell, $\nu$ =+1, ± 2 and +3, as well as a SC state on the hole doped side of $\nu = -2$, with a critical temperature $T_c \sim$ 3.5 K (see SI). Overall the device shows a phase diagram which is very similar to previous reports of hBN non-aligned MATBG devices [19,20,22,23], which is confirmed by examining the crystallographic edges (see SI).

In order to create a JJ in the device, we control both $(V_{BG}, V_{TG})$ to tune the $n$ to $n_{sc}$ = -1.72×10$^{12}$ cm$^{-2}$, where the SC state is at optimal doping. By further changing $(V_{BG}, V_{TG})$ following the relation $\Delta V_{BG} = -(C_{TG}/C_{BG})\Delta V_{TG}$, $n$ can be kept constant, while the carrier density in the junction $n_J$ is continuously tuned (see SI for detailed dual-gate maps). This allows to tune the junction region from a metallic (N) to a SC and into CI state. As the length of the junction is in the order of magnitude of the coherence length of the SC $d_j \sim \xi \sim$ 100 nm (see SI), it is possible to proximitize the junction and to create a Josephson junction[35].

Fig. 1d shows the differential longitudinal resistance of $dV_{xx}/dI$ vs source drain current $I$ and as a function of $n_J$, for a range of fillings which are centered around the SC state $-3 < \nu < -2$. The upper panel shows the corresponding $R_{xx}$ vs. $n_J$ measurement which demonstrates the density ranges of the N, SC and CI states. For $n_J = n_{sc}$, the device is uniformly in the SC state, and forms a SC/SC/SC junction, with a critical current of $I_c >$ 200 nA. However, when $n_J$ is tuned away from this point a SC/SC'/SC junction is created, where we observe a second set of coherence peaks with reduced $I_c$. For density ranges close to $\nu = -3$, -2.2×10$^{12}$ cm$^{-2} \lesssim n_J \lesssim$ -1.86×10$^{12}$ cm$^{-2}$, and close to $\nu = -2$, -1.58×10$^{12}$ cm$^{-2} \gtrsim n_J \gtrsim$ -1.5×10$^{12}$ cm$^{-2}$, the junction region is not intrinsically superconducting, but we still observe a supercurrent, which hints at the creation of a JJ. While close to $\nu = -3$ the junction is metallic and a SC/N/SC is formed, close to $\nu = -2$ the junction is in the correlated insulator state SC/CI/SC. Beyond these density ranges we do not observe a supercurrent, however distinct superconducting non-linearities remain, which are in line with Andreev reflections at the SC interfaces[36].

We further analyze the various gate-induced junctions by applying an out-of-plane magnetic field $B$, where Fig 1e shows the color maps of $dV_{xx}/dI$ vs. $I$ vs. $B$. For the uniform SC/SC/SC junction (center) we observe a diamond-shaped dependence, which is symmetric with the inversion of the current $I_c^+(B^+) = I_c^-(B^+)$ and the $B$-field directions $I_c^+(B^+) = I_c^+(B^-)$, where $I_c^+(I_c^-)$ and $B^+(B^-)$ correspond to the positive (negative) critical current and field. This behavior is in-line with previous reports on SC states in MATBG [19,20,22–24], with a critical magnetic field $B_{c2} \sim 120$ mT. The absence of Fraunhofer oscillations confirms the uniformity of the junction and the absence of a JJ. For both, the SC/N/SC (left) and SC/CI/SC (right) junctions (taken at $n_J$ as marked in Fig. 1d) we observe clear Fraunhofer oscillations with oscillation periods that are consistent with expectations for one flux quantum through the junction (see SI), which unambiguously prove the formation of a gate defined JJ[34,37]. Here the SC/N/SC JJ displays a typical Fraunhofer pattern, which obeys the same symmetries along the $I$ and $B$ directions as the pristine SC state. In stark contrast to this, the SC/CI/SC JJ shows a very unusual Fraunhofer pattern, which is not symmetric with inversion of the current $I_c^+(B^+) \neq I_c^-(B^+)$ and the $B$-field directions $I_c^+(B^+) \neq I_c^+(B^-)$, which indicates time-reversal symmetry breaking. These asymmetries are well seen in the line cuts in Fig. 1 f and g which show $I_c^+$ vs. $B$ and $dVxx/dI$ vs. $I$ measurements respectively. Most strikingly, for both measurements we observe a hysteresis as a function of $B$-field direction.

To understand the unconventional Josephson effect of the SC/CI/SC JJ, we first examine the Fraunhofer pattern with the parameters of Fig. 1g at an elevated temperature of $T = 800$ mK, where the hysteretic behavior at weak magnetic field has not yet developed (Fig. 2a-c). In this regime, the Fraunhofer pattern has the following highly unconventional features: 1. The central peak of the Fraunhofer pattern is shifted to from $B = 0$ to a value of $B \sim 2.5$mT; 2. The Fraunhofer pattern is highly asymmetric with respect to the central peak; 3. The critical current $I_c^+$ does not vanish as a function of $B$; 4. Even more strikingly, when the current direction is reversed, a different Fraunhofer pattern is observed, and the central peak is shifted as shown in Fig.2f. At $B = 0$, for example, the critical current is dramatically different for currents flowing in opposite directions. 5. The critical current shows a hysteresis and the directional dependence of the critical current appears only when the system is pre-magnetized by an external magnetic field larger than a coercive field of $\sim 300$mT (purple and orange lines in Fig.2c). As none of these features are observed in the SC/SC/SC and the SC/N/SC junctions, we suggest that the CI state in the middle of the JJ is an unconventional insulating state responsible for the observed Fraunhofer pattern. Qualitatively, the shift of the central peak, the breaking of the time-reversal symmetry condition $I_c^+(B^+) = I_c^+(B^-)$ and the hysteresis behavior all suggest that time-reversal symmetry is broken and that there is a spontaneous net magnetic flux which is responsible for the shift in the position of the central peak away from $B = 0$.

It is important to note that the observed unconventional Fraunhofer patterns are highly reproducible. The question is: Which microscopic state of MATBG near $\nu = -2$ can explain these? We show below that all the observed experimental features are consistent with the assumption that the CI is an interaction induced valley polarized state with net orbital magnetization.

Specifically, we model the moiré Chern bands of twisted bilayer graphene based on an effective tight-binding model[49,50] with additional $C_2T$ breaking terms to gap out the massless Dirac points. Then, we construct a MATBG based JJ model by assuming the CI in the central region to be a valley polarized state with net Chern number $C = -2$ at filling factor $\nu = -2$. The superconducting part of the JJ is assumed to be a fully gapped superconductor with s-wave pairing (see SI for the details). The Josephson current through this junction is evaluated with the lattice recursive Green's function method (see SI for details). Fig. 2a shows the calculated Fraunhofer pattern ($I_c$ vs. $B$) which displays remarkable agreement with the experiment without fine tuning of theoretical details. Not only does the theory reproduce the oscillation period and the phase shift, but also the asymmetry with respect to the central peak of the unconventional Fraunhofer pattern. Unlike in the case of a conventional Fraunhofer pattern, it is asymmetric with respect to the $B$-field direction, where the $I_c$ of positive $B$ is larger than for negative $B$. We found that removing the $C_2T$ breaking terms will make the bands topologically trivial with no Berry curvatures nor net orbital magnetic moments. In this case, a standard Fraunhofer pattern is obtained (see SI). Therefore, the CI state is likely to be a valley polarized state with topologically nontrivial bands and net orbital magnetization. This suggests that this behavior is a direct consequence of the electronic ground state near $\nu = -2$ carrying orbital magnetization.

This state has been previously identified at slightly elevated magnetic fields $B > 300$ mT [23], which is in good agreement with the observed coercive field of the JJ. It is possible that this state also exists in zero-field, but is obscured due to domain wall formation. Moreover, the orbital magnetic moment of this state is huge ~ 6 $\mu_B$ (Bohr magneton)[47] and produces an out-of-plane magnetic field of $B \sim 3$ mT (see SI for derivation). This is consistent with the experimentally obtained phase shift of $\Delta B \sim 2.5$mT. The phase shift of the Fraunhofer pattern survives up to the critical temperature of the JJ of $T_c \sim 1$ K, and is comparable to the Curie temperature of previously observed orbital magnetic states in hBN aligned[25,26] and non-aligned MATBG[20,23] as well as in twisted mono-bi graphene [38,48].

To explain the directional dependence of the critical current in Fig. 2f, one extra assumption is needed. Namely, that the current $I \sim 10$ nA can induce orbital magnetization switching similar to the current induced orbital magnetization switching, which is observed at a filling of $\nu = 3$ in MATBG [25,26,38, 39 26,41,51]. In other words, a small current can overcome the free energy barrier between two degenerate orbital magnetization states of the CI as schematically illustrated in Fig. 2h. With this assumption, the directional dependence of the critical current is well explained as depicted in Fig. 2f. However, further theoretical study is needed to understand the current induced orbital magnetization switching in this $C = -2$ state.

The Fraunhofer pattern in low temperatures is even more intriguing. Fig. 3a and b show it for $T = 500$ mK, where it is measured by sweeping the $B$-field up (a) and down (b). Strikingly, both Fraunhofer patterns show a phase jump (marked by an arrow), which was not observed at higher temperatures. Comparing the two Fraunhofer patterns, one notices that they are phase shifted, and overall symmetric with respect to the reversal of the current and $B$-field directions, where $I^+(B^+) \sim I_c^-(B^-)$. Its phase jump is hysteretic and occurs at different $B$-fields for the up-and down-sweeps. Such *B*-field hysteresis is better seen in the line-cuts in Fig. 3c, which shows the $I_c^+(B)$ for both field sweeping directions at *T* = 800, 500 and 35 mK. Here we define $\Delta B$ as the difference between the maxima of the $I_c^+(B^+)$ and $I_c^+(B^-)$ sweeps, which follows a critical temperature dependence $\Delta B \sim (1 - T/T_c)^\alpha$ (Fig. 3e) with a Curie temperature $T_c \sim$750mK and $\alpha \sim 0.5$.

Such hysteresis of $I_c(B)$ is a prominent characteristic of ferromagnetic JJs [3–5], which is induced by a switching of the spin orientation. While the *I-B* asymmetry which is indicative of orbital magnetism, continues to be present in the Fraunhofer pattern, the hysteretic features cannot be fully explained by it. These appear almost at an order of magnitude lower temperature and require a two orders of magnitude lower switching field $|B_M| \geq 3$ mT, than observed for the valley polarized state. We hence propose that the state which gives rise to all of these features is a valley and spin polarized state as shown in Fig. 2d. Here the valley polarized state has an order of magnitude larger gap than the spin polarization and hence manifest at elevated temperatures, while the spin polarization only develops at much lower temperatures. The spin texture of the $\nu = -2$ state has been previously discussed experimentally[22,43] and has been recently theoretically predicted to have a net spin polarization[44]. By including both valley and spin polarization into the model Hamiltonian (SI), we can reproduce all the main characteristics of the experimental Fraunhofer pattern (Fig. 3e). The calculated $I_c$ display both the hysteresis and coercive fields behavior. The order of magnitude of the signal and $\Delta B$ also matches the experimental results.

A direct consequence of the remnant magnetization and its current induced magnetization switching in the MATBG JJ, is its non-reciprocal transport. This is demonstrated in the $dV_{xx}/dI$ vs. *I* curves taken at *B* = 0, which show highly non-symmetric behavior with respect to the current direction. As can be clearly seen in Fig. 4a, for a fixed current value $|I| \sim 10 - 50$ nA the device can be superconducting in one current direction, while highly resistive in the other. This behavior enables the creation of a superconducting diode, which is the superconducting analog of a p-n junction, and is highly sought after as a building block for superconducting electronics. Since the magnetization direction can be switched by a small field $B_M$ (red and blue lines in Fig. 4a), the polarity of the current asymmetry can be switched, and the direction of the diode reversed, making it so programmable. We demonstrate the superconducting diode behavior in Fig. 4b, where we apply $|I| \sim 25$ nA and continuously switch the current direction. Simultaneous measurements of the device resistance show that the device is clearly resistive in one current direction and superconducting in the other, depending on the magnetization direction. While the superconducting diode effect has been previously realized in Rashba superconductors due to the magnetochiral anisotropy effect [52,53], as well as in van der Waals JJs[54], the present device offers new capabilities as it is fully tunable and can be operated in zero magnetic field.

To summarize – the zero-field coexistence and gate tunability of the magnetic and topological phases with superconductors in MATBG presents a remarkable opportunity to electronically hybridize these phases through engineering of complex gate induced junctions, This will lead to the creation of ever more complex quantum phases based on the MATBG platform. Also, the so created JJs can shed new light on the underlying ground states of MATBG, as the JJ probe much smaller areas than traditional transport experiments, and are extremely sensitive to magnetic fields.

## Methods:

<u>Device fabrication</u>: The MATBG devices are fabricated using a cut and stack technique. All flakes were first exfoliated in a Si/SiO$_2$ (285 nm) substrate and later picked up using a polycarbonate (PC)/polydimethylsiloxane (PDMS) stamp. All the layers were picked up at a temperature of ~100ºC. The graphene is initially cut with an AFM tip, to avoid strain during the pick-up process. The PC/PDMS stamp was used to pick up first the top graphite layer, the first hBN and the first graphene layer. Before picking up the second graphene layer, the stage is rotated by an angle of 1.1-1.2º. Finally, the bottom hBN and bottom graphite gates were picked up. The finalized stack is dropped on a Si/SiO$_2$ substrate by melting the PC at 180 ºC. The resulting stack is etched into a Hall bar with CHF$_3$/O$_2$ and a 1D contact is formed by evaporating Cr (5 nm)/Au (50 nm). The narrow channel of ~150 nm in the top gate is etched with O$_2$. Before etching the top gate, the device was characterized at $T$ = 35 mK to identify the pair of contacts closest to the magic angle ($\theta$ ~1.1°). The junction was made in between this pair of contacts.

<u>Measurements</u>: Transport measurements were carried out in a dilution refrigerator (Bluefors SD250) with a base temperature of 20 mK. Standard low-frequency lock-in techniques (Stanford Research SR860 amplifiers) were used to measure $R_{xx}$ with an excitation current of 10 nA at a frequency of 13.11 Hz. For the $dV_{xx}/dI$ measurements the excitation current was reduced to 1 nA. The d. c. bias current was applied through a 1/100 divider and a 1 MΩ resistor before combining it with the a.c. excitation. Keithley 2400 Source-meters were used to control the gates as well as serve as the source for the DC current. The measured $dV_{xx}/dI$ signals were filtered and amplified by voltage-preamplifiers SR560 before entering the lock-in amplifiers.

<u>Twist angle extraction</u>: The twist angle is extracted from the phase diagrams shown in Suppl. Fig. S2. The carrier density corresponding to a fully filled superlattice unit cell is extracted to be $n_s$ = (2.88 ± 0.1) ×10$^{12}$ cm$^{-2}$. By applying the relation $n_s = 8\theta^2/\sqrt{3}a^2$, where $a$ = 0.246 nm is the graphene lattice constant, we extract a twist angle $\theta$ = 1.11° ± 0.02°.

## Acknowledgements:
We are grateful for fruitful discussions with Allan MacDonald and Andrei Bernevig. D.K.E. acknowledges support from the Ministry of Economy and Competitiveness of Spain through the "Severo Ochoa" program for Centres of Excellence in R&D (SE5-0522), Fundació Privada Cellex, Fundació Privada Mir-Puig, the Generalitat de Catalunya through the CERCA program, funding from the European Research Council (ERC) under the European Union's Horizon 2020 research and innovation programme (grant agreement No. 852927)" and the La Caixa Foundation. K.T. L. acknowledges the support of the Ministry of Science and Technology of China and the HKRGC through grants MOST20SC04, C6025-19G, 16310219, 16309718 and 16310520. J.D.M. acknowledges support from the INPhINIT 'la Caixa' Foundation (ID 100010434) fellowship programme (LCF/BQ/DI19/11730021).

## Author contributions:
D.K.E and X.L. conceived and designed the experiments; J.D.M., A.D.C. and X.L. fabricated the devices and performed the measurements; J.D.M., A.D.C., S.Y.Y., D.K.E., Y.M.X, X.G. and K.T.L. analyzed the data; Y.M.X, X.J.G. and K.T.L. performed the theoretical analysis; T.T. and K.W. contributed materials; D.K.E. supported the experiments: J.D.M., A.D.C., S.Y.Y., D.K.E., Y.M.X. and K.T.L. wrote the paper.

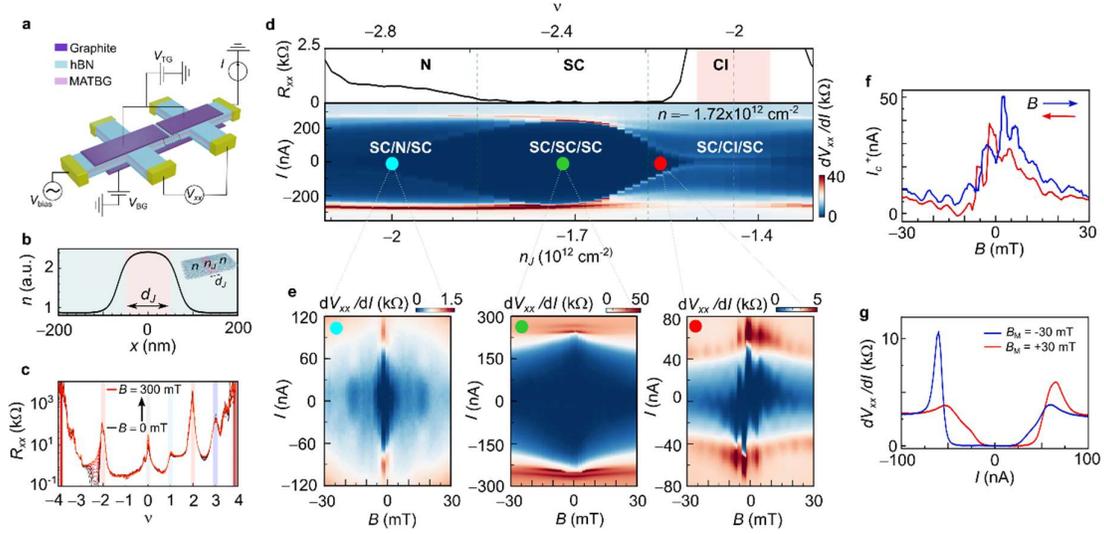

**Fig. 1. Gate tunable JJ in MATBG. a,** Schematic of the measured device and measuring circuit. The top graphite gates are separated by 150 nm. **b,** Electrostatic simulation profile of carrier density $n$ vs. position $x$ setting $n$ and $n_J$ at different values. The inset shows a schematic of the MATBG JJ with two distinct regions created by the gating structure. **c,** Four terminal longitudinal resistance $R_{xx}$ vs. filling factor $\nu$ at different out-of-plane magnetic fields $B$ from 0 mT (black curve) to 300 mT (red curve). **d,** (Top) Magnification of **c** around the SC state $-3 < \nu < -2$. (Bottom) $dV_{xx}/dI$ vs. $I$ at different $n_J$, keeping $n = -1.72\times10^{12}$ cm$^{-2}$ in the SC state. Dashed green vertical lines mark the position where $n_J$ is no longer in the SC state. **e,** Fraunhofer patterns measured at (left) $n_J = -2\times10^{12}$ cm$^{-2}$ (SC/N/SC), (center) $-1.72\times10^{12}$ cm$^{-2}$ (SC/SC/SC) and (right) $-1.56\times10^{12}$ cm$^{-2}$ (close to SC/CI/SC), respectively. The color dots show the corresponding $n_J$ positions in the $dV_{xx}/dI$ vs. $I$ map in **d** bottom. **f,** Positive critical current $I_c^+$ vs. $B$ with $B$ sweeping up (blue) and down (red). **g,** $dV_{xx}/dI$ vs. $I$ at $B = 0$ mT after applying a pre-magnetizing field $B_M = +30$ and $-30$ mT for the red and blue curve. All data in the figure is taken at 35 mK.

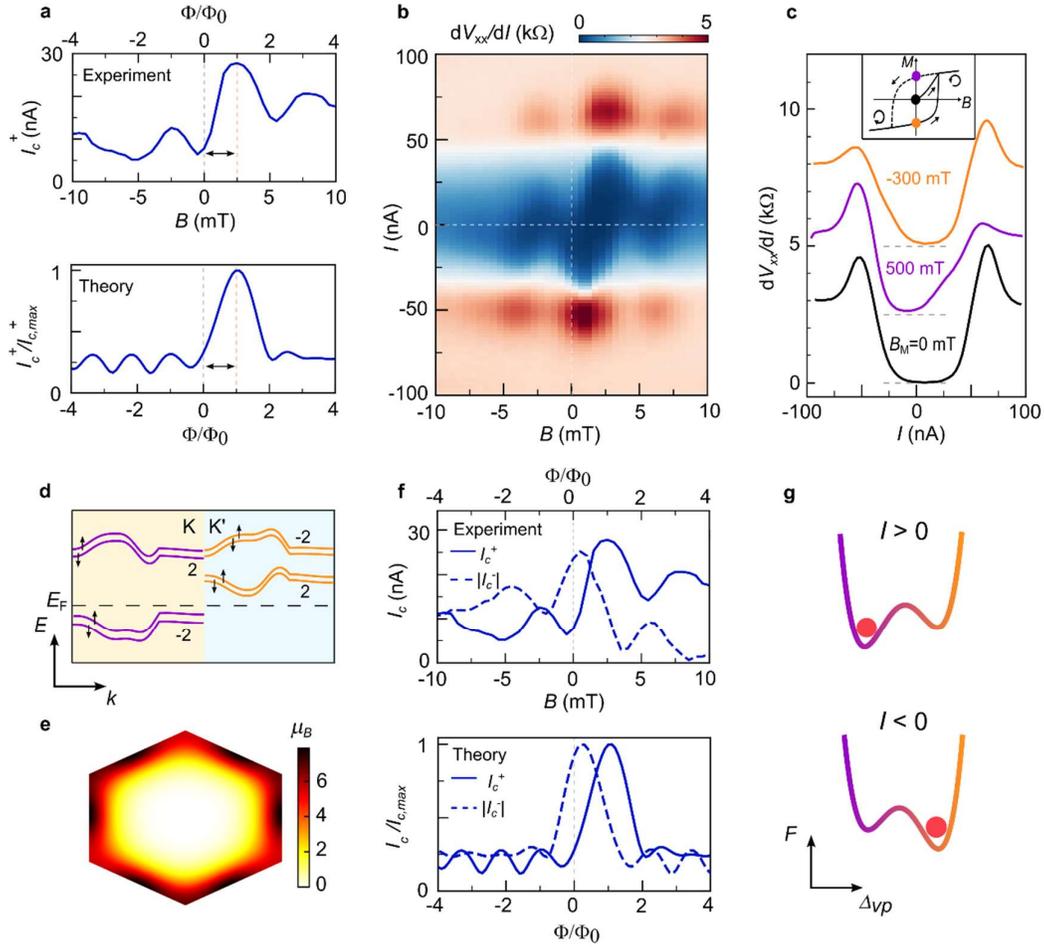

**Fig. 2. JJ with orbital magnetism. a,** (Top) Positive critical current $I_c^+$ vs. $B$ at 800 mK. The vertical dashed lines remark the shift of the $I_c^+$ maximum from zero field. (Bottom) Theoretical $I_c^+$ vs. magnetic flux ($\Phi$) calculated for a MATBG JJ with a valley-polarized $\nu = -2$ state as the weak link. The pattern has been shifted by $+\Phi_0$ to compare with the experiment. **b,** Fraunhofer pattern with $n_J = -1.56\times10^{12}$ cm$^{-2}$ measured at 800 mK. The white dashed lines mark the 0 current and 0 field positions. **c.** $dV_{xx}/dI$ vs. $I$ measured at $B = 0$ mT and $T = 800$ mK right after cooldown (black) and after the sample has been subjected to two opposing pre-magnetizing fields $B_M$. The curves are vertically shifted by 2.5 k$\Omega$ each for clarity. The inset shows a schematic of magnetization $M$ vs. $B$. The colored dots correspond to the magnetic states in which the different $dV_{xx}/dI$ vs. $I$ curves were taken. **d,** Band structure schematic of a valley polarized $\nu = -2$ state, as a function of energy $E$ and momentum $k$, where K and K' mark the two valleys and $E_F$ the position of the Fermi energy. **e,** Magnetic moment distribution in $k$-space in units of Bohr magneton $\mu_B$ within a MATBG unit cell with valley polarization. **f,** (Top) Experimental $I_c^+$ and $|I_c^-|$ vs. $B$, extracted from **b**. Reversing the current direction inverts the line-shape of the curve and changes the shift in magnetic field. (Bottom) Theoretical $I_c^+$ and $|I_c^-|$ vs. $\Phi$ for a MATBG JJ with a valley polarized $\nu = -2$ state as the weak link. To compare with the experiment, a shift of $+\Phi_0$ and $+0.2\Phi_0$ was added to $I_c^+$ and $|I_c^-|$, respectively. **g.** Schematic illustration of controlling the valley polarization via a DC current.

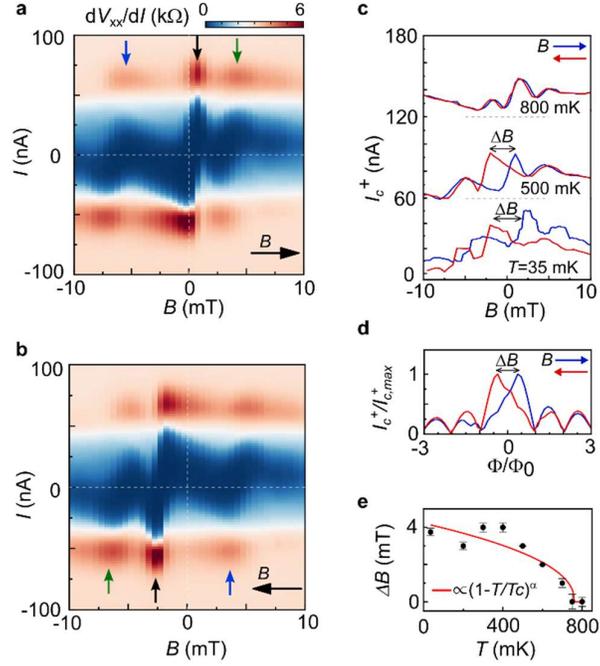

**Fig 3. Evolution of magnetic hysteresis with temperature. a-b,** Fraunhofer patterns measured at 500 mK with field sweeping up (**a**) and down (**b**). The white dashed lines mark the 0 current and 0 field positions. The colored arrows highlight a period change in the pattern and the fact that by rotating **a** by 180º one would get the periodicity of **b**. **c,** $I_c^+$ extracted from the Fraunhofer patterns with the magnetic field sweeping up (blue) and down (red) at 800, 500 and 35 mK. The curves are vertically shifted by 60 nA each for clarity. **d,** Theoretically calculated $I_c^+$ vs. $\Phi$ for a spin polarized MATBG JJ, with the field sweeping up (blue) and down (red). **e,** Extracted $\Delta B$ vs. $T$ for $I_c^+$. The red curve is a fit to $(1 - T/T_c)^\alpha$ with fitting parameters $T_c \approx 750$ mK and $\alpha \approx 0.45$.

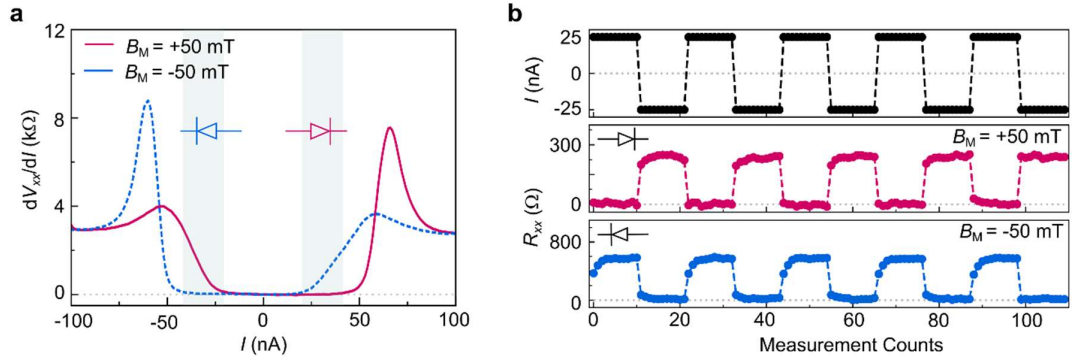

**Fig. 4. Zero-field switchable superconducting Josephson diode. a,** $dV_{xx}/dI$ vs. $I$ measured setting $n_{J=}$ -1.56×10$^{12}$ cm$^{-2}$ at 35 mK. All curves are taken at $B$=0 mT after pre-magnetizing the sample at $B_M$ = +50 mT (red) or $B_M$ = -50 mT (blue). The shaded gray regions mark the values of current at which the diode behavior is observed. **b,** Switching between resistive and superconducting state by changing the direction of $I$ as shown in the top panel. By applying opposite $B_M$, the diode behavior is inverted (red and blue curves).

# Supplementary Information for 'Magnetic Josephson Junctions and Superconducting Diodes in Magic Angle Twisted Bilayer Graphene'

J. Díez-Mérida[1], A. Díez-Carlón[1], S. Y. Yang[1], Y.-M. Xie[2], X.-J. Gao[2], K. Watanabe[3], T. Taniguchi[3], X. Lu[1], K. T. Law[2] and D. K. Efetov[1]*

1. ICFO - Institut de Ciencies Fotoniques, The Barcelona Institute of Science and Technology, Castelldefels, Barcelona, 08860, Spain
2. Department of Physics, Hong Kong University of Science and Technology, Clear Water Bay, Hong Kong, China
3. National Institute for Materials Science, 1-1 Namiki, Tsukuba, 305-0044, Japan

*E-mail: dmitri.efetov@icfo.eu

## A. Fabrication and transport characterization of the MATBG device.

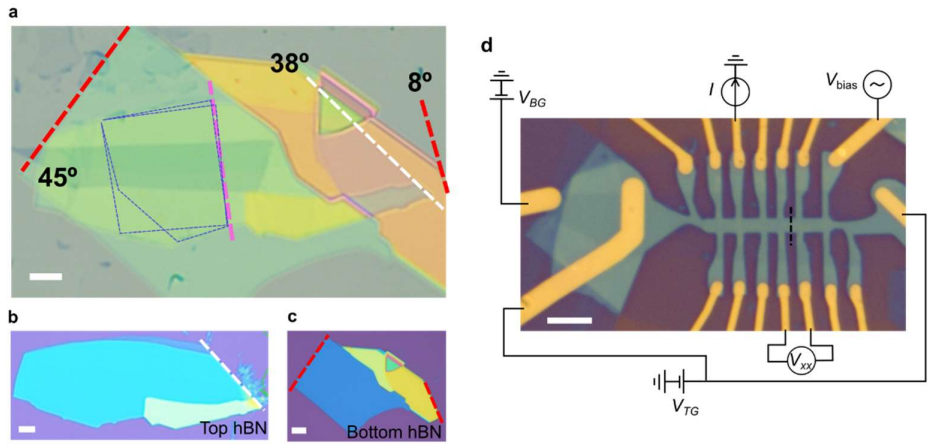

**Figure S1. Optical image of the device and alignment to hBN. a,** Resulting final stack, where the red and white dotted lines show misalignment between the edges of top and bottom hBN and those of the two graphene flakes composing the MATBG in blue. **b – c,** Optical micrographs of top and bottom hBN used for stacking the van-der-Waals heterostructure shown in **a**. The white and red dotted lines mark the crystallographic edges of the two hBN flakes. **d**, Optical picture of the final device, with a schematic of our 4-probe measurement setup. The black dotted line marks the position of the narrow-etched region of the top-graphite gate, corresponding to the location of the weak link in the junction. Scale bars equal 5 μm in all figures.

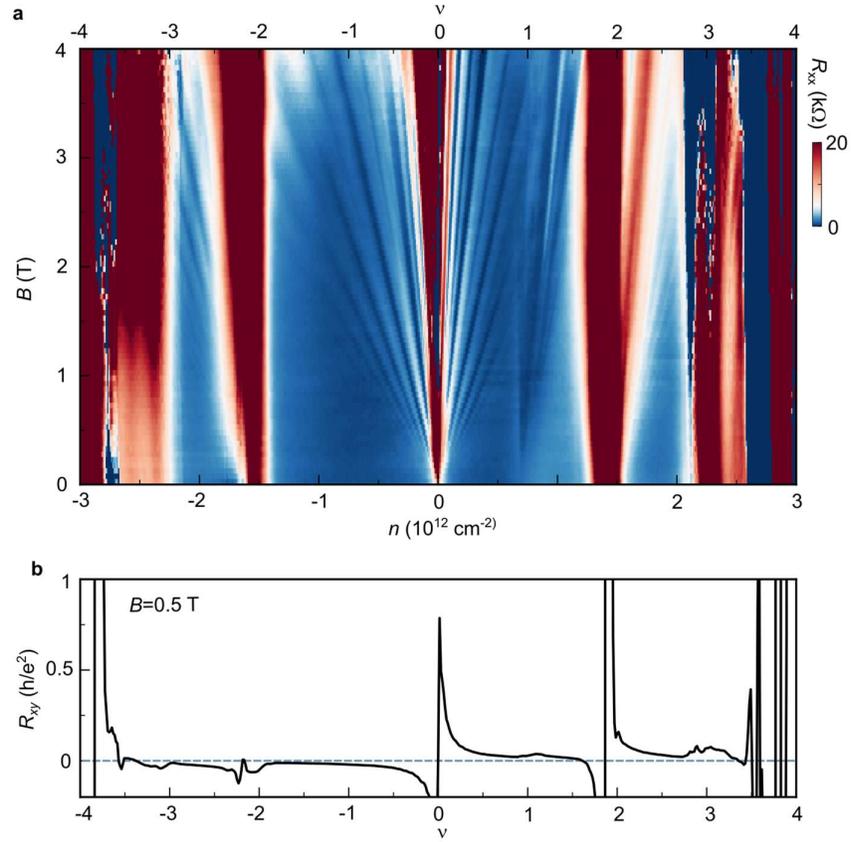

**Figure S2. Landau fan diagram and absence of quantized plateau Hall resistance at low field. a,** Longitudinal resistance $R_{xx}$ vs. carrier density $n$ and perpendicular magnetic field $B$, where Landau levels were used to calculate the twist angle of the device. **b,** Anti-symmetrized transverse resistance $R_{xy}$ vs. $n$ at $B = 0.5$ T, plotted in units of $h/e^2$, where $h$ is Planck's constant and $e$ is the electron charge. No quantized plateau Hall resistance close to $\nu = -2$ is observed. All data in the figure is taken setting $V_{TG} = 0$ V.

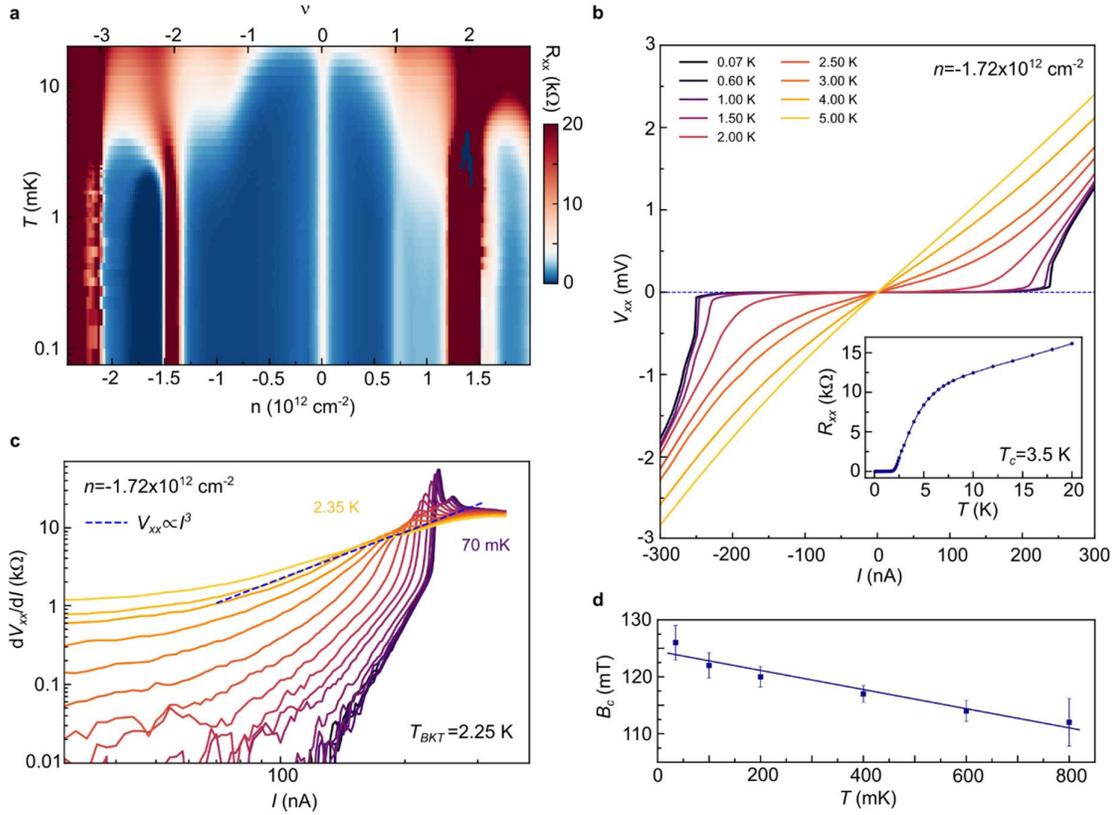

**Figure S3. Temperature dependence and characterization of the superconducting state.**
**a,** $R_{xx}$ as a function of $n$ and temperature $T$. **b,** Current-voltage ($I$-$V$) characteristic of the superconducting state at $n = -1.72 \times 10^{12}$ cm$^{-2}$ for a wide range of temperatures. From the lowest temperature data, we extract a critical current of $I_c = 240$ nA. The inset shows $R_{xx}$ as a function of $T$, where a transition to the superconducting state is observed with a critical temperature $T_c = 3.5$ K, taken as the temperature at which $R_{xx}$ equals to 50% of the normal state resistance. **c,** Differential resistance $dV_{xx}/dI$ vs. d.c. current $I$ at various temperatures. Fitting $V_{xx} \propto I^3$ yields a BKT transition temperature $T_{BKT} = 2.25$ K. **d,** Perpendicular critical field $B_c$ vs $T$ taken as half of the normal state resistance values. From linearly fitting $B_c = (\Phi_0/2\pi\xi_{GL}^2)(1 - T/T_c)$, we extract a Ginzburg-Landau superconducting coherence length $\xi_{GL} = 106 \pm 8$ nm at $T = 0$, where $\Phi_0 = h/2e$ is the superconducting flux quantum. All data in the figure is taken setting ($V_{TG} = 0$ V).

## B. Dual gate characterization of the MATBG JJ device and electrostatic model.

Fig. S4a shows $R_{xx}$ as a function of back gate voltage $V_{BG}$ and top gates voltage $V_{TG}$, measured at 1.8 K. This dual-gate map contains two distinct features, corresponding to the $n$ and $n_J$ regions sketched in Fig. 1c of the main text. The diagonal features, marked as gray dashed lines in the figure, are those gated both by the back gate and the top gates. Its carrier density is given by $n = C_{BG}V_{BG} + C_{TG}V_{TG}$, where $C_{BG}$ and $C_{TG}$ are the capacitances of the back gate and top gates, respectively. The features corresponding to the junction are displayed as green dashed lines. This region is mainly gated by the back gate, but having a small contribution from the top gates due to the effect of stray fields[1]. This explains why these features are not perfectly vertical but rather have a finite slope. The carrier density of this region follows $n_J = C_{BG}V_{BG} + \alpha C_{TG}V_{TG}$, where $\alpha \ll 1$. As such, we can independently gate the two regions.

Fig. S4b is a zoom-in of the dual-gate map measured at 35 mK, where the dark blue region corresponds to the superconducting state. By following this dark blue region parallel to the grey lines, the dual-gated main regions remain in the superconducting phase, whereas the single-gated junction region can be continuously changed in doping, creating different configurations of a JJ, as indicated by the white squares. This diagonal line is the one followed to take the $dV_{xx}/dI$ colormap of Fig. 1d in the main text.

In order to gain insight on how the electrostatics of the dual gated architecture of our device works, we have performed electrostatic simulations by solving the Poisson equation with a finite difference method[2]. The generalized Poisson equation is given by $\nabla \cdot [\epsilon(\mathbf{r}) \nabla V(\mathbf{r})] = \rho(\mathbf{r})/\epsilon_0$, where $\epsilon$ is the dielectric constant ($\epsilon = 4$ for hBN), $V$ is the electrostatic potential, $\rho$ is the density of electric charges and $\epsilon_0$ is the vacuum permittivity. The problem is solved self-consistently with an iterative approach following a successive over-relaxation method[3]. Dirichlet boundary conditions (BCs) are used for the electrodes: $V(z = 20$ nm$) = V_{TG}$ and $V(z = -20$ nm$) = V_{BG}$. The sides and top regions with no gate are set with Neumann BCs, satisfying $\partial V_x = 0$ and $\partial V_y = 0$, respectively. The model is strictly electrostatic, not including any band structure properties of the MATBG. Fig. S5a shows the response of the case where the weak link is set very close to the $\nu = -2$ CI (SCS configuration). The linecut shown in Fig. S5b is extracted from the dashed gray line of the figure, where the MATBG would be located. Here the electrostatic potential is converted to $n$ by a factor related to the capacitances of the top and back gates. From the simulations it is possible to estimate the effective size of the JJ ($d_J$), which does not match the length of the etched region, $d \approx 150$ nm. Since $n$ has a slow transition from the SC to the CI state, the effective length of the JJ is $d_J \approx 100$ nm.

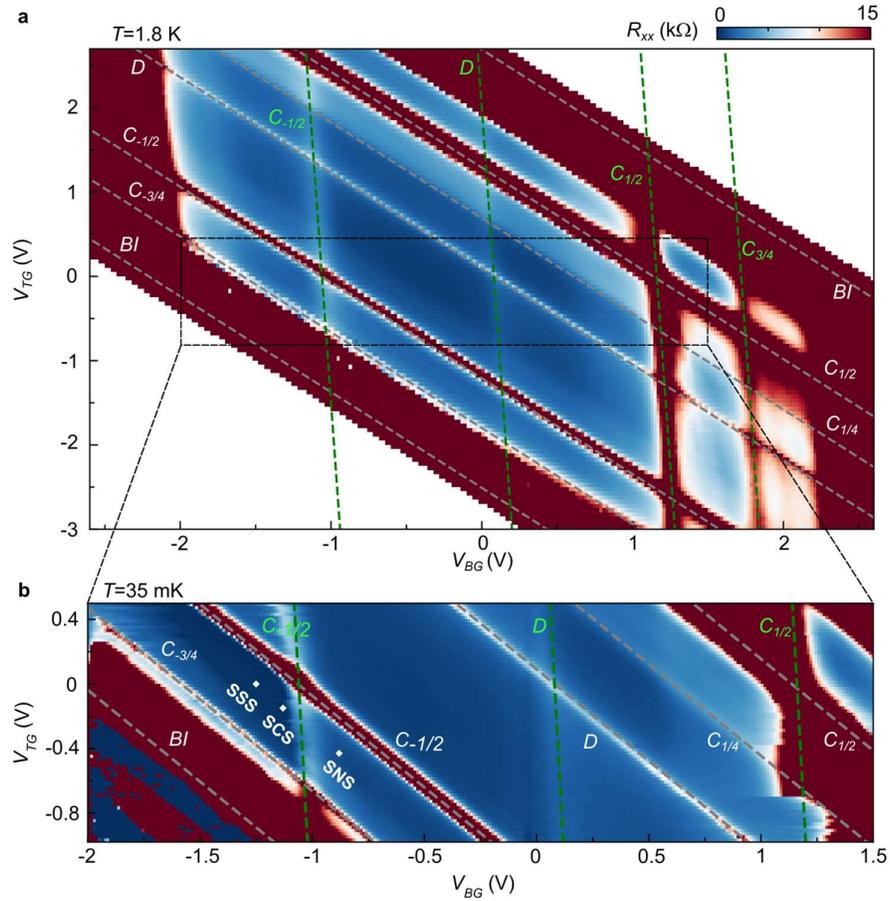

**Figure S4. Dual-gated maps. a,** $R_{xx}$ vs. $V_{BG}$ and $V_{TG}$ at $T = 1.8$ K. The diagonal features marked with grey dashed lines are the integer fillings of the main regions ($n$ in Fig. 1b of main text), gated by both the back gate and top gates. Slightly tilted vertical features, fitted with green dotted lines, are the integer fillings of the junction region ($n_J$ in Fig. 1b of main text), mainly gated by the back gate. Labels of the integer filling factors marked by dashed lines correspond to: band insulator (*BI*) at full filling, correlated insulator at plus/minus three-quarter filling ($C_{\pm 3/4}$), plus/minus half filling ($C_{\pm 1/2}$) and one quarter filling ($C_{1/4}$) and Dirac point (*D*). **b,** Zoom-in of black-delimited area in **a** taken at base temperature $T = 35$ mK, where the superconducting state is fully developed. White squares mark the diagonal line in the map at which both main regions are kept in the superconducting state, and the junction region is set at different doping.

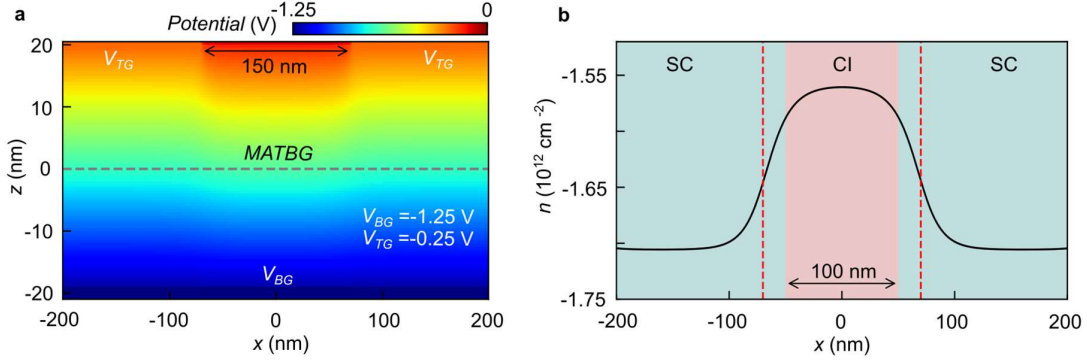

**Figure S5. Simulations of the electrostatic potentials governing our devices. a,** Calculated potentials as a function of position $x$ and height $z$, for a junction configuration with $V_{BG} = -1.25$ V and $V_{TG} = -0.25$ V. The back gate and top gates are placed at $z = -20$ nm and $z = 20$ nm, respectively. The etched region is marked by the double arrow at the top of the figure. The dashed gray line shows the position at which the MATBG would be located. **(b)** Extracted linecut of the electrostatic potential from **a** at $z = 0$ nm, the MATBG position. Vertical red dashed lines show the position where the top graphite was etched. The light green and red regions show the extension of the SC and CI states. The effective length of the JJ in this case is 100 nm, as marked in the figure.

### C. Evolution of the asymmetric Fraunhofer patterns with $n$ and $T$, and $I_c$ extraction

Fig. S6a displays an extension of the $dV_{xx}/dI$ vs $I$ map shown in Fig. 1e of the main text, to better identify the $n_J$ positions of the Fraunhofer patterns, which are labelled with color dots. Fig. S6b-e show the evolution of the Fraunhofer patterns when the weak link is being doped at different carrier densities $n_J$. Upon tuning $n_J$ from $-1.72\times10^{12}$ cm$^{-2}$ (SSS configuration) to $-1.56\times10^{12}$ cm$^{-2}$ (SCS configuration), the Fraunhofer evolves from being symmetric in both current and field towards being asymmetric. When $n_J \gtrsim -1.35\times10^{12}$ cm$^{-2}$ or $n_J \lesssim -2.1\times10^{12}$ cm$^{-2}$, the coherence is lost in our JJ.

To study the behavior of the JJ with magnetic field, a detailed analysis of $I_c$ at different currents and fields is performed in the main text (Fig. 2 and 3). The exact procedure by which we define and extract $I_c$ is shown in Fig. S7a. We define $I_c$ as the $I$ value at which the resistance is non-zero anymore. We define a $R$ threshold value of $\approx 100$ Ω, displayed as the dashed red line and $I_c$ is taken as the $I$ value at which the $dV_{xx}/dI$ exceeds such a threshold, marked by the red cross in the figure. Since this procedure is highly dependent on the value of the defined threshold, several values are considered, and an error is taken by calculating the standard deviation of all the extracted $I_c$ values. The error obtained from this analysis is the error which is plotted in the temperature dependence of $\Delta B$ in Fig. 3. The main value of $I_c$ is chosen as the best fit to the contour in the Fraunhofer pattern, as shown in Fig. S7b. In this example we plot the 35 mK data shown in Fig. 1h of the main text. The extracted $I_c^+$ and $I_c^-$ are plotted as a red and green dashed line over the map. The same procedure is followed for all the Fraunhofer patterns mentioned in the main text.

Finally, in this section, we discuss the temperature dependence of the Fraunhofer pattern of the SCS configuration and its magnetic hysteresis. Fig. S8 shows the SC/CI/SC Fraunhofer patterns measured at different temperatures that are not shown in the main text. For every temperature, the Fraunhofer is measured with field sweeping in both directions. The insets of each of these plots demonstrate the fact that the position and magnitude of coherence peaks is reversed at zero field for opposite field sweeping directions, due to the hysteretic behavior inside the JJ. The hysteresis is further evidenced in the $I_c$ vs. $B$ plots (Fig. S8 l-p) where $|I_c^+(B_{up})| \neq |I_c^+(B_{down})|$, and is gradually lost until $T = 800$ mK, where $I_c^+(B_{up}) |=| I_c^+(B_{down}) |$. The temperature dependence of $\Delta B$ shown in Fig. 3f of the main text is extracted from this data.

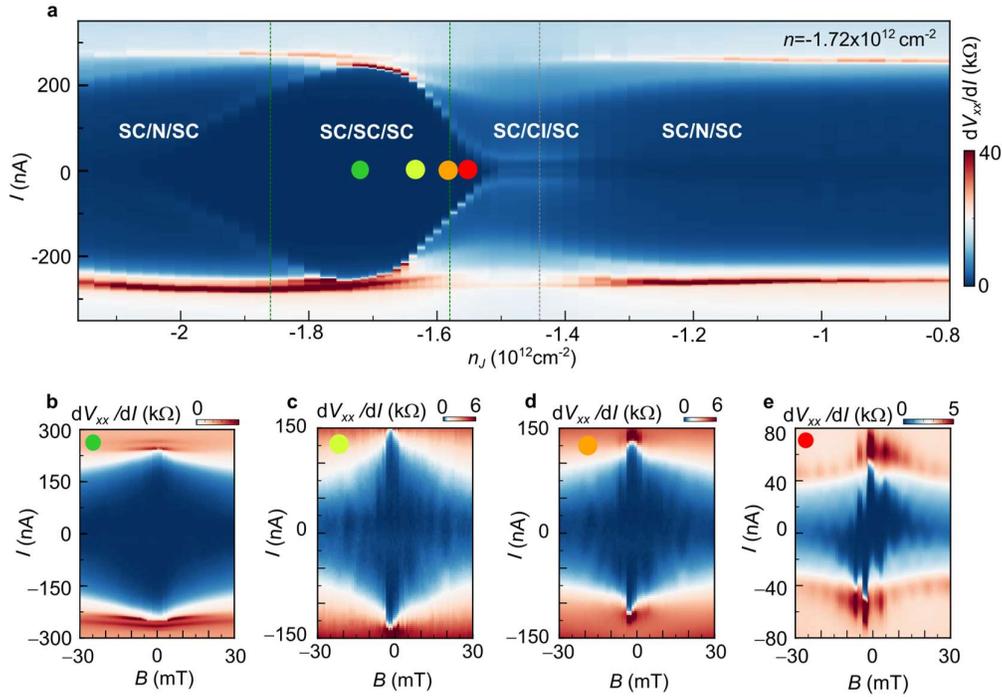

**Figure S6. Carrier density dependence of the asymmetry in the Fraunhofer patterns. a**, $dV_{xx}/dI$ vs. $I$ and $n_J$ at $T = 35$ mK, keeping $n$ in the superconducting state. The dashed lines are the same as in Fig. 1e of main text. **b-e,** Fraunhofer patterns with the weak link being at different doping $n_J$, corresponding to those labelled by the color circles in **a**. The Fraunhofer patterns evolve from being symmetric in the SC/SC/SC to being asymmetric in the SC/CI/SC configuration.

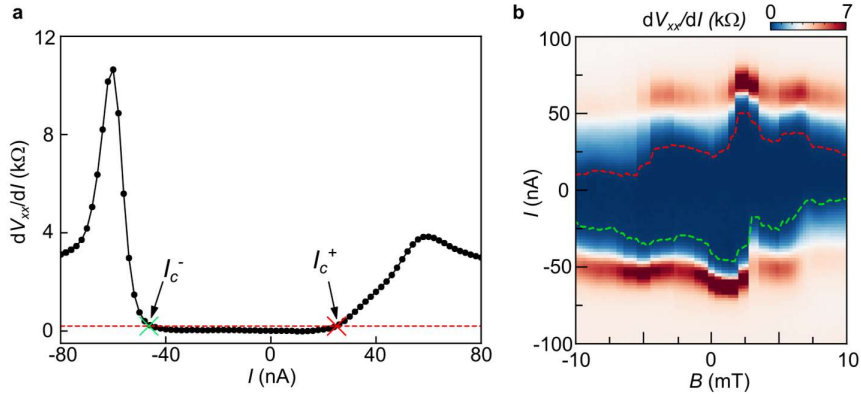

**Figure S7. $I_c$ extraction from the $dV_{xx}/dI$ vs. $I$ data.** The plotted data corresponds to the same Fraunhofer pattern as in Fig. 1h of the main text. **a,** $dV_{xx}/dI$ vs. $I$ linecut of **b** at zero field. The red dashed line marks a threshold value defining the end of the superconducting state. **(b)** $dV_{xx}/dI$ vs. $I$ and $B$. The extracted $I_c^+$ and $I_c^-$ are sketched as dashed red and green lines.

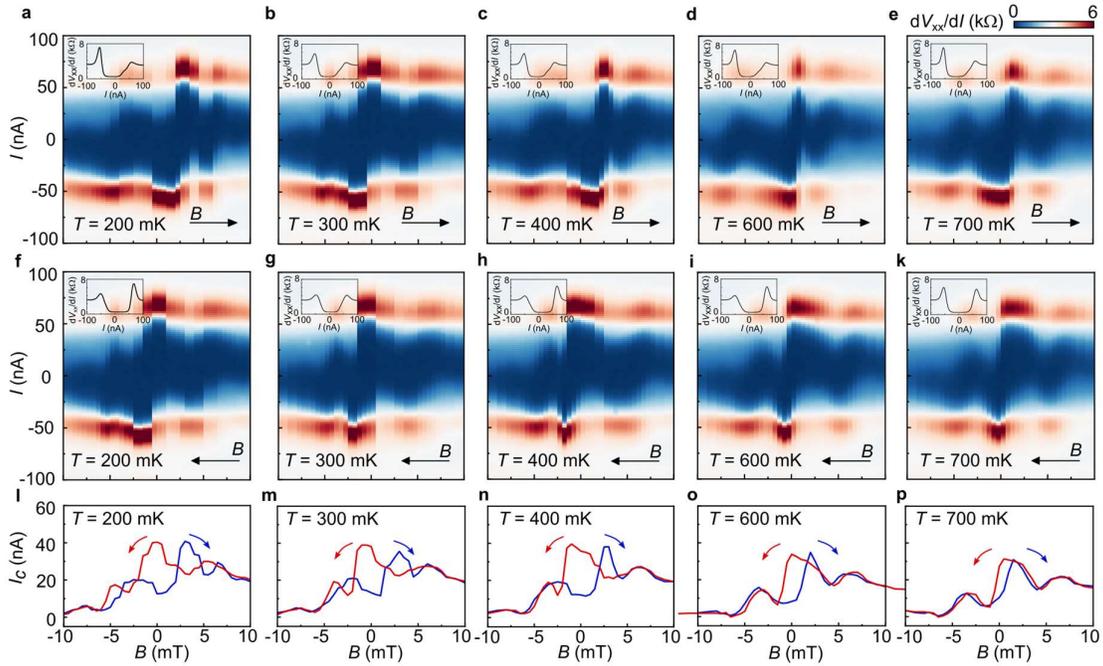

**Figure S8. Temperature dependence of the SCS Fraunhofer pattern and its hysteresis. a-e,** Fraunhofer patterns measured at $n_J = -1.56 \times 10^{12}$ cm$^{-2}$, at increasing temperatures from 35 mK to 700 mK and by sweeping the magnetic field up. Insets of each plot show a $dV_{xx}/dI$ vs. $I$ linecut at $B=0$ T. **f-k,** Corresponding Fraunhofer patterns to **a-e** respectively, at the same temperature but opposite magnetic field sweep direction. **l-p,** Extracted positive critical current $I_c^+$ from the corresponding temperatures and with field sweeping up (blue) or down (red).

## D. Two-dimensional planar and three-dimensional bulk behaviour of the Fraunhofer patterns.

In a two-dimensional superconductor, the film thickness is smaller than the London penetration depth $\lambda$. In this case, the magnetic flux can penetrate into the superconductor and the spatial distribution of magnetic field is governed by the Pearl length[4] $\Lambda = 2\lambda^2/t$, where $t$ is the film thickness. It has been previously shown [5,6] that the critical current of a two-dimensional JJ under a perpendicular magnetic follows the relation $\Delta B_{2D} \approx 1.8\Phi_0/w^2$, where $w$ is the lateral size of the junction. This is different from a three-dimensional bulk Josephson Junction[7], where the period of the oscillations follows $\Delta B_{3D} \approx \Phi_0/wd$, where $d$ is the length of the junction. According to our JJ geometry ($w \approx 1.2 \pm 0.1$ μm and $d_J \approx 100$ nm), $\Delta B_{3D}$ is estimated to be ~ 16±1 mT, which is more than three times greater than $\Delta B_{2D}$ ~ 2.5±0.5 mT, as shown in the calculations in Fig. S9.

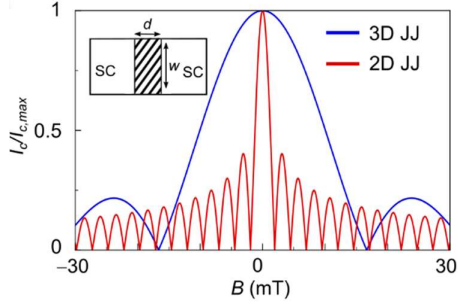

**Figure S9. 2D planar and 3D bulk Fraunhofer patterns.** Calculations of the expected critical current for a JJ of our sample dimensions ($w \approx 1.2$ μm and $d_J \approx 100$ nm) in the 2D planar regime and the 3D bulk regime.

## E. The model Hamiltonian for the TBG based-Josephson junction

In this section, we present the model Hamiltonians for this Josephson junction, which will be used to calculate the Fraunhofer patterns. The Josephson junction studied in this experiment is rather special in the sense that all parts are formed by the MATBG only. As schematically plot in Fig. S10a, this Josephson junction consists of three parts: the left and right parts are the superconducting MATBG which conducts supercurrents, and the middle part that links the two superconducting parts is also formed by a MATBG but being closed to half-filling, where correlated insulating phases are known to be established.

To capture the bands of MATBG, we used a hexagonal lattice model[8,9], which is written as:

$$H_0 = \sum_{j,\boldsymbol{\delta}^{(1)}} (c_j^\dagger \cdot \hat{T}_1 \cdot c_{j+\boldsymbol{\delta}^{(1)}} + \text{H.c.}) + \sum_{j,\boldsymbol{\delta}^{(3)}} (c_j^\dagger \cdot \hat{T}_3 \cdot c_{j+\boldsymbol{\delta}^{(3)}} + \text{H.c.}). \quad (1)$$

where the $c_j = (c_{j+,\uparrow}, c_{j+,\downarrow}, c_{j-,\uparrow}, c_{j-,\downarrow})$ denotes a four-component electron annihilation operator with $c_{j\pm}$ being the annihilation operator for $p_x \pm ip_y$ orbitals, ↑/↓ representing spin up/down at a site $j$, and the first-nearest neighbor and the fifth-nearest neighbor (to break the emergent SU(4) symmetry) hopping terms are considered with hopping matrix $\hat{T}_1 = [t_1, 0; 0, t_1] \otimes \sigma_0$, $\hat{T}_3 = [t_2, 0; 0, t_2^*] \otimes \sigma_0$, the connecting vectors $\boldsymbol{\delta}^{(1)} = \{\boldsymbol{\delta}_1, C_3\boldsymbol{\delta}_1, C_3^2\boldsymbol{\delta}_1\}$, $\boldsymbol{\delta}^{(3)} = \{\boldsymbol{\delta}_3, C_3\boldsymbol{\delta}_3, C_3^2\boldsymbol{\delta}_3\}$ (see Fig.S11b), $\sigma_0$ is an identity matrix defined in spin-space. This model respects $D_3$ point group symmetry, spin $SU(2)$ symmetry, and the orbital $U(1)$ symmetry. It is also worth noting that we can regard the two orbitals $p_x + i\xi p_y$ as a representation of two valleys $\xi = \pm$ [8]. In the calculations, we adopted the model parameters from ref. [8], where $t_1 = 0.331$ meV, $t_2 = -0.01 + 0.097i$ meV.

Based on the insight about the presence of the spin and orbital magnetism, we write the model Hamiltonian of this Josephson junction formed by the MATBG as:

$$H = H_0 + H_h + H_{mu} + H_{order}. \quad (2)$$

where $H_0$ captures the bands of MATBG as written in Eq. (1). Let us introduce the details for other terms as follows.

To enable the bands to form orbital Chern bands, we added the Haldane Hamiltonian[10]:

$$H_h = \sum_{j \in A, \delta^{(2)}} [(c_j^\dagger \cdot \hat{T}_2 \cdot c_{j+\delta^{(2)}} + \text{H.c.}) + m_{stag} c_j^\dagger \cdot c_j] + \sum_{j \in B, \delta^{(2)}} [c_j^\dagger \cdot \hat{T}_2^\dagger \cdot c_{j+\delta^{(2)}} + \text{H.c.} - m_{stag} c_j^\dagger \cdot c_j], \quad (3)$$

where the next-nearest connecting vectors $\delta^{(2)} = (\delta_2, C_3\delta_2, C_3^2\delta_2)$, $A$ and $B$ are the sublattice indices, the complex hopping matrix $\hat{T}_2 = [t_f e^{i\varphi}, 0; 0, t_f e^{-i\varphi}] \otimes \sigma_0$ and $m_{stag}$ is a staggered potential. Without loss of generality, we set $t_f = 0.2t_1$, $m_{stag} = 0.1t_1$, $\varphi = \pi/2$, where the bands are topological [3]. This consideration is motived by another experiment with the same experimental setup, where orbital moiré Chern bands were found[11]. Noted what is esstential in our discussions is that the bands are orbital Chern bands carrying net orbital magnetism. In this sense, a phenomenological term as Eq. (3) is sufficient for our purpose. Physically, these terms can be induced by the combination of the moiré potential and some $C_2T$ breaking terms. Note although we added the Haldane Hamiltonian for the whole MATBG, it is only crucial for the middle part of the junction to generate the orbital magnetism upon valley polarization. And we checked whether the Haldane Hamiltonian is added or not in the left and right superconducting part will not affect the features of the Fraunhofer pattern as long as the chemical potential is far from the charge neutrality point.

The filling difference between different regions is captured by $H_\mu$:

$$H_\mu = -\sum_{j \in L,R} \mu_{sc} c_j^\dagger \cdot c_j - \sum_{j \in M} \mu c_j^\dagger \cdot c_j \quad (4)$$

with $\mu_s, \mu$ respectively, denotes the chemical potential of the superconducting MATBG parts in the left, right part (label as **L**, **R**) of the junction, and the MATBG near half-filling in the middle part (label as **M**) of the junction (Fig. S11a). In the calculation, we fix $\mu_s = -0.5$ meV in the superconducting part which is around the middle of moiré bands.

Finally, to capture the phase orders in different parts, we introduced a mean-field Hamiltonian $H_{order}$, which is written as:

$$H_{order} = \sum_{j \in L,R} (|\Delta_{sc}| e^{i\phi_{L(R)}} c_j^\dagger (\tau_x \otimes i\sigma_y) c_j^\dagger + \text{H.c.}) + \sum_{j \in M} c_j^\dagger (\Delta_{sp} \tau_0 \otimes \sigma_z + \Delta_{vp} \tau_z \otimes \sigma_0) c_j. \quad (5)$$

where Pauli matrices $\tau, \sigma$ operates on orbital-, spin-space, respectively, $\Delta_{sc}$, $\Delta_{sp}$ and $\Delta_{vp}$ denotes the order parameters of superconducting states, spin-polarized states, valley-polarized states, respectively. The first term captures the superconducting phase $\phi_{L(R)}$ of the left (right) part of the junction. Here, we assumed the superconducting order parameter takes the conventional BCS form, which is most likely to be stabilized by the electron-phonon interaction[12,13]. Considering that the key features of the Fraunhofer patterns arise from the spin and orbital magnetism in the middle part, similar Fraunhofer patterns may be found even when the paring is unconventional, such as $p$-wave or $d$-wave paring, although some details such as the amplitude of supercurrent should be different. A detailed study of the junction in the presence of both unconventional pairings and magnetism is beyond the scope of this work, but in general this is interesting and we leave it as future work. The second term captures the spin- and valley-polarized states that generate the spin and orbital magnetism in the middle part of the junction.

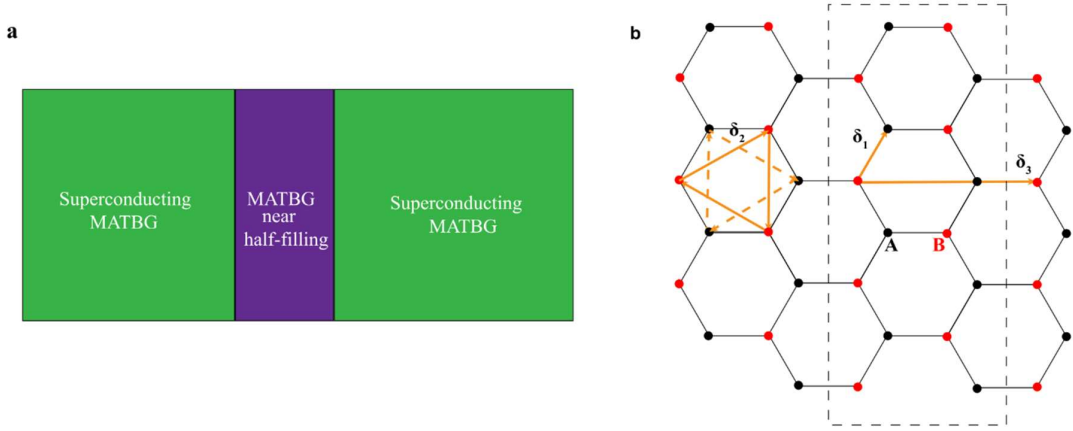

**Figure S10. Sample and lattice geometry considered in the theoretical calculation for this MATBG based-Josephson junction. a,** A schematic plot of the studied twisted bilayer graphene based Josephson junction. **b,** A figure to illustrate the effective hexagonal lattice for the Fraunhofer pattern calculations, where $\delta_j$ are bond vectors, $A$ and $B$ label two different sub-lattices, the yellow arrow indicates there is a hopping between two sites. The dash lines highlight the shape of the recursive unit in our Fraunhofer pattern calculations.

### F. The method for the calculation of the Fraunhofer patterns

After introducing the model Hamiltonian for the Josephson junction formed by the MATBG, we now present the detailed process for the calculation of Fraunhofer patterns. To evaluate the Fraunhofer patterns, we need to take account into the effects of external magnetic fields **B** on the junction, where we consider **B** is finite only in the middle part. This is done by replacing the hopping terms in the middle part with the Peierls substitution as:

$$t_{jj'} \to t_{jj'} e^{i\frac{e}{\hbar}\int \mathbf{A}(\mathbf{r})\cdot d\mathbf{r}}. \tag{6}$$

The Landau gauge fields are taken as $\mathbf{A}(\mathbf{r}) = (0, Bx, 0)$, where we set $x$-direction to be the transverse direction of the junction. Let us insert $\mathbf{A}(\mathbf{r})$ back to Eq. (6) and we can obtain

$$t_{jj'}e^{i\frac{e}{\hbar}\int A(\mathbf{r})\cdot d\mathbf{r}} = t_{jj'}e^{i\frac{2\pi\bar{x}\Delta y}{\sqrt{3}a^2 N_t}\frac{\Phi}{\Phi_0}}. \tag{7}$$

Here, $a$ is the lattice constant, $N_t$ is the total number of hexagonal plaquettes, the flux quantum $\Phi_0 = h/2e$, the total flux $\Phi = \frac{3}{2}N_t B a^2$, $\bar{x} = (x_1 + x_2)/2$, $\Delta y = y_2 - y_1$ with $\mathbf{j} = (x_1, y_1)$, $\mathbf{j'} = (x_2, y_2)$.

After this Peierls substitution, we can evaluate the Josephson current using the lattice recursive Green's function method[14,15]. Here, the recursive unit is shown in Fig. S11b (within dashed lines), which is formed by two zigzag chains. The specific recursive processes are:

(i) obtain the edge Green's function of the superconducting MATBG in the right and left part of Josephson junction by recursively calculating:

$$G_{ll}^{L(R)}(\omega_n) = (i\omega_n - H_{ll}^{L(R)} - \Sigma_{l,l}^{L(R)}(\omega_n))^{-1}, \tag{8}$$

$$\Sigma_{l,l}^{L} = V_{l,l-1}G_{l-1,l-1}(\omega_n)V_{l,l-1}^{\dagger}, \quad \Sigma_{l,l}^{R} = V_{l,l+1}G_{l+1,l+1}(\omega_n)V_{l,l+1}^{\dagger}. \tag{9}$$

where $l$ is the index for the recursive unit, $\omega_n = (2n+1)\pi T$ is the Matsubara frequency with $T$ as temperature. Here the intra-unit Green's function $G_{ll}(\omega_n)$ and self-energy $\Sigma_{l,l}$ is evaluated iteratively until $G_{ll}(\omega_n)$ is saturated using the intra-unit Hamiltonian $H_{ll}^{L(R)}$ of the left and right part, the hopping Hamiltonian $V_{l+1,l}$ between the unit $l$ and $l+1$. Note that $H_{ll}^{L(R)}$ and $V_{l+1,l}$ are Bogoliubov de Gennes (BdG) Hamiltonians defined in Nambu basis $(c_j, c_j^{\dagger})$. For example, $V_{l+1,l} = \text{diag}[V_{l+1,l}^{ee}; V_{l+1,l}^{hh}]$ with $V_{l+1,l}^{hh} = -V_{l+1,l}^{ee*}$. Also note the initial Green's function $G_{00}(\omega_n) = (i\omega_n - H_{ll}^{L(R)})^{-1}$, and the superconducting phase in $H_{ll}^L$ and $H_{ll}^R$ are $\phi_L$ and $\phi_R$, respectively. After these iterations, we can obtain the intra-unit Green's function $G_{l_1,l_1}(\omega_n)$ and $G_{l_3,l_3}(\omega_n)$, where $l_1$ ($l_3$) is the index of the first (last) unit of the middle part.

(ii) obtain the intra-unit Green's function $G_{l_2 l_2}(\omega_n)$, nearest-unit Green's function $G_{l_2,l_2+1}(\omega_n)$, where $l_2$ labels a unit cell in the middle part. To obtain this, we can first perform a similar recursive calculation as step (i) for the $l_1$ to $l_2 - 1$ and for $l_3$ to $l_2 + 1$ part, i.e.,

$$G_{ll}(\omega_n) = (i\omega_n - H_{ll}^{M} - \Sigma_{ll}(\omega_n))^{-1}, \tag{10}$$

$$\Sigma_{l+1,l+1} = V_{l+1,l}G_{ll}(\omega_n)V_{l+1,l}^{\dagger}. \tag{11}$$

Here initial Green's function for $l_1$ to $l_2 - 1$ and $l_3$ to $l_2 + 1$ part are $G_{l_1,l_1}^L$ and $G_{l_3,l_3}^R$. In this way, we can obtain $\Sigma_{l_2-1,l_2-1}(\omega_n)$ and $\Sigma_{l_2+1,l_2+1}(\omega_n)$. Then we can obtain

$$G_{l_2,l_2}(\omega_n) = (i\omega_n - H_{ll}^{M} - \Sigma_{l_2-1,l_2-1}(\omega_n) - \Sigma_{l_2+1,l_2+1}(\omega_n))^{-1}, \tag{12}$$

$$G_{l_2,l_2+1}(\omega_n) = G_{l_2+1,l_2+1}(\omega_n)V_{l_2,l_2+1}^{\dagger}G_{l_2,l_2}(\omega_n), \tag{13}$$

$$G_{l_2+1,l_2}(\omega_n) = G_{l_2,l_2}(\omega_n)V_{l_2,l_2+1}G_{l_2+1,l_2+1}(\omega_n). \tag{14}$$

(iii) evaluate the Josephson current:

$$I(\Delta\phi) = -iT\sum_n \text{Tr}[\tilde{V}_{l_2,l_2+1}G_{l_2+1,l_2}(\omega_n) - \tilde{V}_{l_2,l_2+1}^{\dagger}G_{l_2,l_2+1}(\omega_n)], \tag{15}$$

where the current depends on the phase difference $\Delta\phi = \phi_L - \phi_R$ and in particular, being distinct from $V_{l+1,l}$, the $\tilde{V}_{l+1,l} = \text{diag}[V_{l+1,l}^{ee}; -V_{l+1,l}^{hh}]$, i.e., there is an additional sign in the hole part, as the electron and hole carry opposite currents. More details of this method can be referred to ref. [14,15]

## G. Orbital Chern bands and the orbital magnetic moment.

In this section, we illustrate the features of energy bands of normal states in our model Hamiltonian. The energy bands along high symmetry lines with and without valley polarization are shown in Fig.S11a and Fig.S11b. Fig.S11a reproduces the bands of MATBG given in ref. [8,9], where the red lines and green lines label the bands from two valleys, respectively. In Fig.S11b, we added a valley polarization $\Delta_{vp} = 1.5$ meV and the Haldane terms, but leaves the spin polarization to be zero so that each band is doubly degenerate. Here the degeneracy at $\overline{K}$ is lifted by the Haldane terms, which break $C_2T$ symmetry, and near half-filling $\nu = -2$, an isolated band from one valley would be occupied as a result of the valley polarization. Since the cooper pairs are pairing states from intervalley, the supercurrent will be weakened stronger with the increase of valley-polarization strength $\Delta_{vp}$.

Another crucial feature is that the bands are topological nontrivial now, i.e. they are orbital Chern bands, due to the Haldane terms (see Eq. 3). To illustrate this, the distribution of Berry curvature in the Brillouin zone is depicted in Fig.S11c. It clearly shows the band carries a finite Chern number $C = 1$. Note the Chern numbers $C$ of Chern bands in each valley carry the same sign but are opposite for opposite valleys (Fig.S11b), as the two valleys are related by the time-reversal operation. As we mentioned, the appearance of orbital Chern bands in these MATBG samples near half-filling have been found in ref. [11] already.

An important consequence of the orbital Chern bands is to give rise to orbital magnetism[16,17]. To show this explicitly here, we calculated the orbital magnetic moment $m_{n\mathbf{k}}$ carried by these orbital Chern bands with[16–19]:

$$m_{n\mathbf{k}} = \frac{e}{2\hbar} \sum_{l \neq n} \text{Im} \frac{<u_{n\mathbf{k}}|\nabla_\mathbf{k} H(\mathbf{k})|u_l(\mathbf{k})> \times <u_{l\mathbf{k}}|\nabla_\mathbf{k} H(\mathbf{k})|u_{n\mathbf{k}}>}{E_{l\mathbf{k}} - E_{n\mathbf{k}}}, \quad (16)$$

where $H(\mathbf{k})|u_{n\mathbf{k}}> = E_{n\mathbf{k}}|u_{n\mathbf{k}}>$, $l, n$ are band indices. To see the magntude of the orbital magnetic moment of per electron for n-th band $M_{n,orbit}$, we can calculate:

$$M_{n,orbit} = \frac{1}{N} \sum_\mathbf{k} m_{n\mathbf{k}}, \quad (17)$$

where $N$ is the number of $k$ points. We found the calculated $M_{n,orbit}$ is about 3 $\mu_B$ for each moire band according to the orbital magnetic mement distribution shown in main text Fig.2e. In the next part, we will show such a large orbital magnetic moment will give rise to a shift in the Fraunhofer pattern being compable to the experiment.

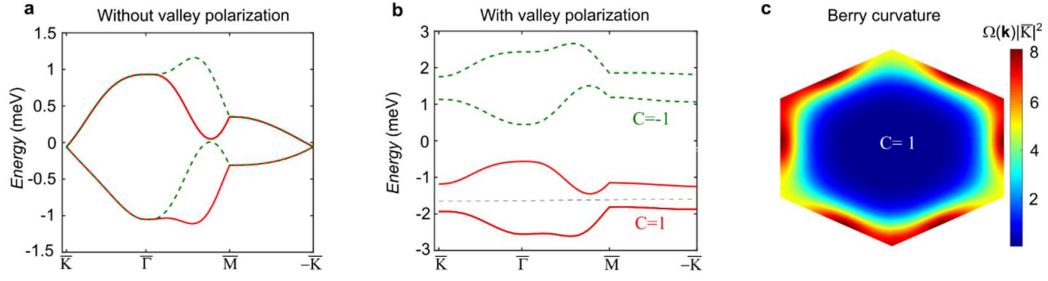

**Figure S11: a-b** Band structures of MATBG with and without valley polarization. The solid red bands are from one valley, while the dashed green bands are from the opposite valley. In **b** the Haldane term is added to lift the degeracies at $\pm \overline{K}$ and a dashed black line is used to indicate the half-filling position, where there is an insulating gap. **c,** Distribution of Berry curvature of the lowest bands in **b** in the Brillouin zone, which clearly shows a Chern number $C = 1$.

## I. Estimate the phase shift in Fraunhofer patterns arising from the presence of net magnetization

The magnetic field in a sample is:

$$B = \mu_0(H + M), \tag{18}$$

where $H$ can be regarded as the applied magnetic fields, $\mu_0 = 4\pi \times 10^{-7}$ T· m/A is the vacuum magnetic permeability, and the magnetization, including the orbital and spin part, is given by:

$$M = M_{orbit} + M_{spin}. \tag{19}$$

Let us assume each electron carries $\gamma u_B$ magnetic moment, where $u_B = e\hbar/2m_e = 9.27 \times 10^{-2}$ J/T, and denote the electron density is $\rho$ m$^{-3}$. The magnetization thus is written as:

$$M = \gamma \rho u_B. \tag{20}$$

Therefore, the effective $B$ field from the magnetization is:

$$B_{eff} = \mu_0 M = \gamma \rho \times 1.13 \times 10^{-2} \text{ T.} \tag{21}$$

For the twisted bilayer graphene near the magic angle, the moire pattern lattice constant is $L_M = 14$nm and the thickness is about 0.34 nm. Near the half-filling $\nu = -2$, the electron density (in unit of $m^{-3}$) is:

$$\rho = 2/(14^2 \times \sqrt{3}/2 \times 0.34) \times 10^{27} \approx 3.47 \times 10^{25} \tag{22}$$

We can also use the experimental carrier density $\sim 1.72 \times 10^{12}$ cm$^{-2}$, and the electron density becomes:

$$\rho \approx 1.72/0.34 \times 10^{25} = 5.07 \times 10^{25}. \tag{23}$$

Using the estimated electron density, we obtain the shift fields due to the presence of magnetism as:

$$B_{eff} \approx 0.4\sim0.6\ \gamma\ \text{mT} \qquad (24)$$

There are two moire bands and each moire band carries a orbital magnetic moment of $3\mu_B$ according to the previous section so that $\gamma \sim 6u_B$ for the valley polarized state at $\nu = -2$. As a result, the shifting field $B_M$ is eastimated to be 2.4~3.6 mT, being consistent with the observed 2.5 mT phase shift in the Fraunhofer pattern. After adding this shifting field, the peak of Fraunhofer pattern will be moved to a finite field, as illustrated in Fig. S12b.

## J. Details for the theoretically calculated Fraunhofer patterns

In this section, we summarize some details for the theoretically calculated Fraunhofer patterns. To obtain the critical current as a function of $\Phi$, i.e., Fraunhofer pattern, at each external field, we vary the phase different over 0 to $2\pi$ and obtain the largest positive and smallest negative current as $I_c^+(\Phi)$ and $I_c^-(\Phi)$. In our calculations, we set pairing gap $\Delta = 0.1$ meV, and we tuned $\mu$, which effectively changes the filling, and the order parameter in the middle region: $\Delta_s$ and $\Delta_{vp}$, which give rise to the spin and orbital magnetism.

First, we set $\mu$ in the middle of the bands and did not add the valley polarization. In this case, the middle region behaves as a normal metal and the whole junction is a SC-N-SC junction. As a result, we obtained the standard Fraunhofer pattern, as shown in Fig.3a, where $I_c \sim |\frac{\sin(\pi\Phi/\Phi_0)}{\pi\Phi/\Phi_0}|$.

Next, we turn on the valley polarization $\Delta_{vp}$. As we discussed, the valley polarization has several effects: (i) As we consider the inter-valley pairing, the valley polarization would suppress the formation of the cooper pairing and reduce the supercurrent through this junction; (ii) The valley polarization will break the time-reversal symmetry and lift the valley degeneracy. Consequently, the combination of orbital Chern bands and valley polarization results in net orbital magnetism. Especially, as we showed, the orbital magnetic moment can be tens of Bohr magneton and leads a shift in Fraunhofer pattern with a shift flux $\Phi_M = B_{eff}S$ (as illustrated in Fig.S13b), $S$ is the junction area. On the other hand, we found the Fraunhofer pattern with the valley polarization $\Delta_{vp}$ typically follows a standard Fraunhofer pattern when $\mu$ is away from half-filling, being expected for an SC-I-SC junction or an SC-N-SC junction (I denotes insulator).

However, we found the Fraunhofer pattern deviates from the standard Fraunhofer pattern when the $\mu$ is closed to the half-filling, and especially, some asymmetric behaviors are established. The calculated Fraunhofer pattern near half-filling with $\Delta_{vp} = 0.8$ meV, $\mu = -0.9$ meV is plotted in Fig.S12c. Here we chose a sizable valley polarization so that the two valleys are fully separated. It can be seen that the Fraunhofer pattern exhibits a similar asymmetric behavior as observed in the experiment. We also checked the asymmetric behaviour is quite robust as long as the $\Delta_{vp}$ can polarize two valleys considerably, as shown in Fig. S12d-i, where we display the Fraunhofer patterns at half-filling for various $\Delta_{vp}$.

However, we found the asymmetry with respect to $B$ fields is lost when the Haldane terms are removed, regardless of the strength of valley polarization (Fig. S12d). This suggested the orbital magnetic moment plays a crucial role in giving rise to the asymmetric behaviour. This understanding is also reasonable as the magnetic fields couple with the orbital moment and can shift the Chern bands. Because this shifting is opposite for opposite fields (the applied fields are small~several mT would not reverse the sign of orbital moment), it thus could result in different critical supercurrent through this junction for opposite fields. On the other hand, the orbital magnetic moment gradually concentrates near half-filling, and thus we obtained a relatively symmetric Fraunhofer pattern in calculation when $\mu$ is artifically tuned away from half-fillings. Therefore, our interpretation that this asymmetry with respect to $B$ in Fraunhofer patterns arises from the coupling between external $B$ fields and orbital magnetic moment of bands is qualitatively consistent with our calculations. This interpretation would also directly imply there is a current-induced magnetization switching in the experiment as discussed in the main text, since the asymmetric behaviour is reversed for opposite current directions.

Beyond the bands level, here we also give a mean-field level discussion on how the coupling between $B$ fields and orbital magnetization can introduce asymmetric Fraunhofer patterns in this junction. We assumed the sign of valley polarization $\Delta_{vp}$ with the switch of current direction and the $B$ fields couples with $\Delta_{vp}$ as this valley polarized state carries orbital magnetic moment. We can describe this switching with a phenomenological free energy as $F_{free} = a\Delta_{vp}^2 + b\Delta_{vp}^4 + \lambda_1 \text{sign}(I)\Delta_{vp} - \lambda_2 B\Delta_{vp}$. As shown in the main text Fig. 2f, the sign of valley polarization depends on the current direction in this case. Moreover, we would like to point out the coupling between magnetic fields and valley polarization order, i.e., $B\Delta_{vp}$ term, in general, gives a larger $\Delta_{vp}$ at the positive $B$ field region, and a smaller $\Delta_{vp}$ at the negative $B$-field region. As a result, this can result in another mechanism for the possible asymmetry in Fraunhofer patterns at positive $B$ field and negative $B$ field range. Note the coercive field $B_c$ ~300 mT is much larger than the applied fields, so the field itself would not switch the valley polarization, and the change of $\Delta_{vp}$ due to the change of $B$ field thus should be quite small. It is worth mentioning that in our calculation, we found a 0.01 meV change in $\Delta_{vp}$ is enough to enable the asymmetry induced by this mechanism to be seen. However, whether several mT $B$ fields used in the experiment can cause a meaningful change in $\Delta_{vp}$, such as large as 0.01 meV, would depend on the specific details, such as domain walls and interactions.

But regardless of different levels of interpretations, our experimental data and our calculations strongly suggest asymmetric symmetry Fraunhofer patterns indicate a net orbital magentization near $\nu = -2$ and it can be switched by the supercurrent. To our knowledge, the orbital magnetic moment-induced asymmetric Fraunhofer pattern has not been realized so far. We thus think a more qualitative study is quite desirable and would leave this as another work.

Finally, we discuss the feature of the Fraunhofer pattern in the presence of spin magnetism. As we discussed, the spin magnetism is likely to build up in the low-temperature range ($T$ <800 mK) according to the measured Fraunhofer patterns. To show explicitly, we calculated the Fraunhofer pattern by adding a spin polarization $\Delta_{sp}$. Specifically, we inputted $\Delta_{sp}$ as a hysteresis loop of $B$ field as shown in Fig.S13a, where we define the saturated spin polarization $\Delta_{sp}$ as $\Delta_{saturate}$. In Fig. S13b and main text Fig.3d, we show the Fraunhofer pattern with $\Delta_{saturate}/\Delta_{sc} = 0.5$ and $\Delta_{saturate}/\Delta_{sc} = 1$, $\Delta_{sc}$ is the superconducting gap. By sweeping the $B$-field from negative to positive and from positive to negative, we also found a hysteresis in the Fraunhofer pattern induced by the hysteresis of spin magnetism. This is due to the fact that the supercurrent carried by singlet Cooper pairings will be reduced by the increase of spin polarization. It thus would be also expected the hysteresis will be more prominent when the size of spin polarization characterized by $\Delta_{saturate}$ is comparable to the superconducting paring gap $\Delta_{sc}$, as we showed in main text Fig.3d.

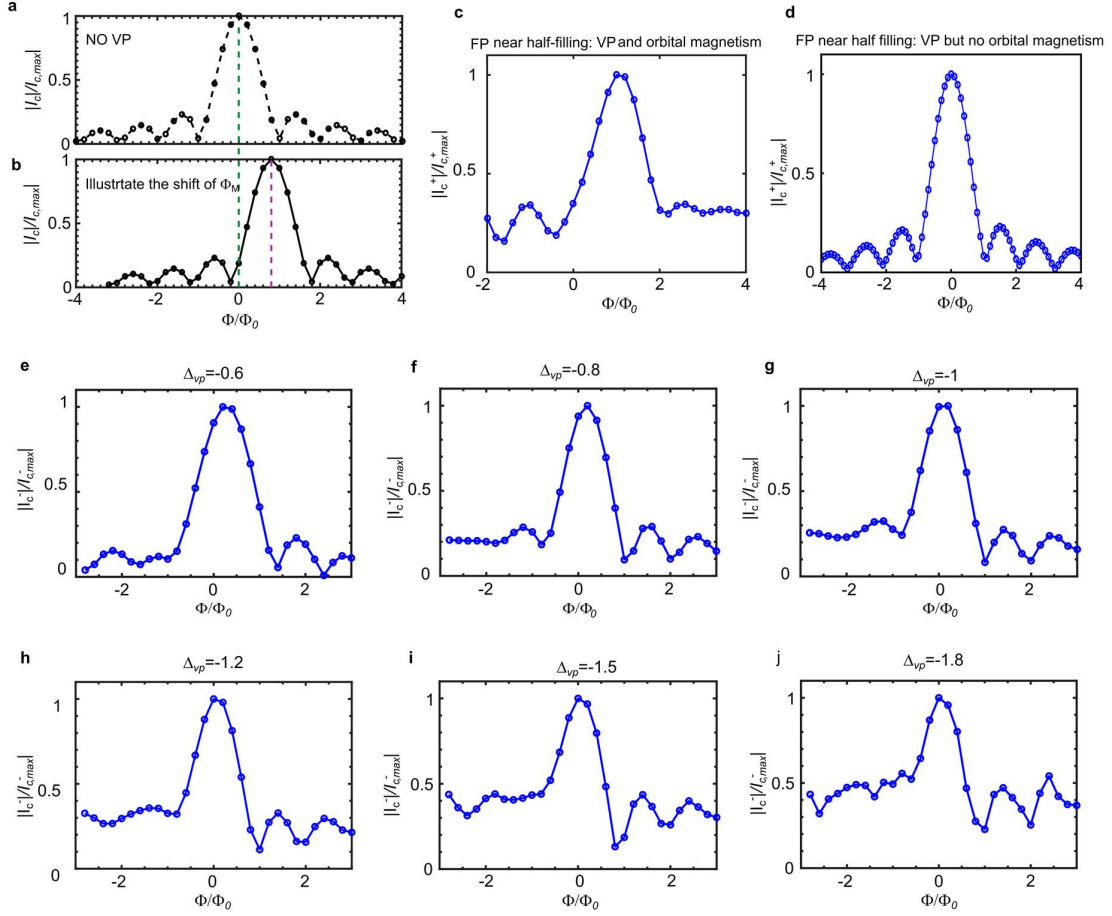

**Figure S12: Fraunhofer pattern in the presence of valley polarization and orbital magnetism**. **a,** Standard Fraunhofer pattern when the middle range is a metal, where $I_c$ denotes the critical current. **b,** Shift of $\Phi_M$ in the Fraunhofer pattern due to the orbital magnetism. **c,** Typical Fraunhofer pattern near half-filling, which supports an insulating gap. The asymmetry induced by the valley polarization is quite clear. **d.** Fraunhofer pattern near half-filling ""with no *C2T* symmetry breaking terms. The asymmetry is not present in this case as there is no orbital magnetization. **e-j,** Calculated Fraunhofer pattern with various valley polarized strength $\Delta_{vp}$ (in units of meV) near half-filling. To compare with the experiment, we also add a shift of $+0.8\Phi_0$ and $+0.2\Phi_0$ for $I_c^+$ and $I_c^-$, respectively, in the Fraunhofer pattern for (c) and (i).

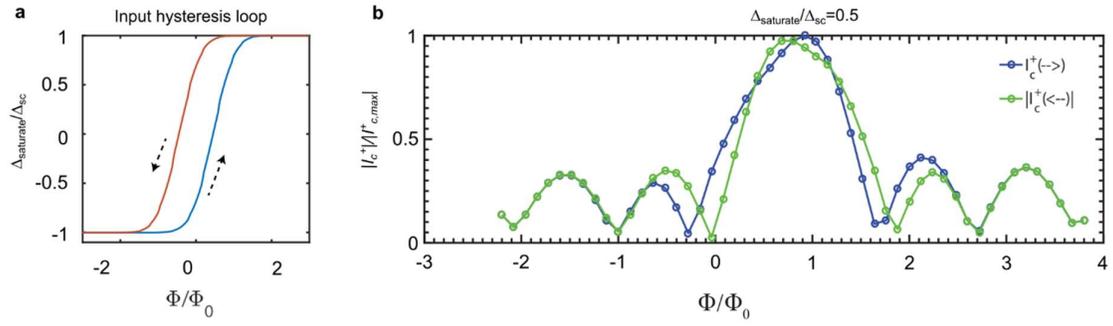

**Figure S13: Hysteresis in Fraunhofer pattern from spin magnetism. a,** Hysteresis loop inputted in the calculation, where $\Delta_{saturate}$ denotes the saturated spin polarization. **b,** Fraunhofer pattern with $\Delta_{saturate}/\Delta_{sc} = 0.5$.